\documentclass[journal]{IEEEtran}
\usepackage{lineno,hyperref}
\usepackage{float}
\usepackage{graphicx}
\usepackage{epstopdf}
\usepackage{algorithm}
\usepackage{algorithmicx}
\usepackage{algpseudocode}
\usepackage{amsmath}
\usepackage{amstext}
\usepackage{amssymb}
\usepackage{caption}
\usepackage{subfigure}
\usepackage{color}
\usepackage{multirow}
\usepackage{subfigure}
\usepackage{listings}
\usepackage{xcolor}
\usepackage[numbers,sort&compress]{natbib}
\usepackage{setspace}

\begin{document}

\title{Blockchain-aided Secure Semantic Communication for AI-Generated Content in Metaverse}

\author{Yijing Lin,
        Hongyang Du,
		    Dusit Niyato,~\IEEEmembership{Fellow,~IEEE}, \\
        Jiangtian Nie,
        Jiayi Zhang,
        Yanyu Cheng,
        and Zhaohui Yang
\thanks{Corresponding author: Hongyang Du.}
\thanks{Yijing Lin is with the State Key Laboratory of Networking and Switching Technology, Beijing University of Posts and Telecommunications, 100876, Beijing, China (e-mail: yjlin@bupt.edu.cn).}
\thanks{Hongyang Du, Dusit Niyato, Jiangtian Nie, and Yanyu Cheng are with Nanyang Technological University, 639798, Singapore (e-mail: hongyang001@e.ntu.edu.sg; dniyato@ntu.edu.sg; jnie001@e.ntu.edu.sg; yanyu.cheng@ntu.edu.sg).}
\thanks{Jiayi Zhang is with School of Electronic and Information Engineering, Beijing Jiaotong University, Beijing 100044, China (jiayizhangg@bjtu.edu.cn).}
\thanks{Zhaohui Yang is with the College of Information Science and Electronic Engineering, Zhejiang University, China (e-mail: yang zhaohui@zju.edu.cn).}
}

\maketitle

\begin{abstract}
The construction of virtual transportation networks requires massive data to be transmitted from edge devices to Virtual Service Providers (VSP) to facilitate circulations between the physical and virtual domains in Metaverse. Leveraging semantic communication for reducing information redundancy, VSPs can receive semantic data from edge devices to provide varied services through advanced techniques, e.g., AI-Generated Content (AIGC), for users to explore digital worlds. But the use of semantic communication raises a security issue because attackers could send malicious semantic data with similar semantic information but different desired content to break Metaverse services and cause wrong output of AIGC. Therefore, in this paper, we first propose a blockchain-aided semantic communication framework for AIGC services in virtual transportation networks to facilitate interactions of the physical and virtual domains among VSPs and edge devices. We illustrate a training-based targeted semantic attack scheme to generate adversarial semantic data by various loss functions. We also design a semantic defense scheme that uses the blockchain and zero-knowledge proofs to tell the difference between the semantic similarities of adversarial and authentic semantic data and to check the authenticity of semantic data transformations. Simulation results show that the proposed defense method can reduce the semantic similarity of the adversarial semantic data and the authentic ones by up to 30\% compared with the attack scheme.
\end{abstract}

\begin{IEEEkeywords}
Metaverse, Blockchain, Semantic Communication, Semantic Attacks, Semantic Defenses
\end{IEEEkeywords}

\maketitle

\section{INTRODUCTION}
\IEEEPARstart{T}{he} word Metaverse was coined in the science-fiction novel Snow Crash \cite{stephenson2003snow} to describe a virtual reality society in which people utilize digital avatars to symbolize themselves and experience the world. In recent years, Metaverse has received attention from academia and industries as a novel Internet application due to the advancement of Augmented Reality (AR), Virtual Reality (VR), Artificial Intelligence (AI), and Blockchain. Specifically, extending reality (AR/VR) and AI technologies can provide users with immersive experiences and facilitate continuous data synchronization from the physical domain to the virtual domain, which is supported by data perceived by edge devices and services driven in Metaverse. Blockchain can help the physical and virtual domains to share information and construct economic systems in a decentralized manner. Thus, the Metaverse can be viewed asthe integration of multiple technologies supported by massive data interactions between the physical and virtual domains.

One of the significant advantages of Metaverse is that people can conduct experiments in the virtual world that cannot be conducted in the real world. Because the virtual world can be created based on real-world data, e.g., images, sensing data, and text, the experimental result in Metaverse can be used to guide the real world. One example is the virtual transportation networks \cite{ng2022stochastic}, i.e., virtual environments in which users can safely train automatic driving algorithms and test vehicles. To build such virtual environments, virtual service providers (VSPs) can leverage network edge devices to capture images from the real world and then render the virtual objects. However, this process entails three challenges as follows:
\begin{enumerate}
    \item[{\bf{D1)}}] How to use the data collected from the real world, e.g., images, to achieve fast virtual world building?
    \item[{\bf{D2)}}] How to improve the efficiency of data transmission to facilitate interactions of the physical and virtual domains?
    \item[{\bf{D3)}}] How to ensure the security of the data received by VSP to ensure that the virtual world can be accurately synchronized with the real world?
\end{enumerate}

Since virtual transportation networks require extensive interactions between the physical domain and Metaverse, edge devices can be utilized to capture images of geographical landmarks. However, converting these captured images into a consistent style for the virtual world is complicated. In several virtual service designs, VSPs need to hire digital painters to pre-process images~\cite{serkova2020digital}, which is time-consuming and costly. The boom in AI has brought alternative solutions. AI-generated content (AIGC) technology~\cite{du2022enabling} allows VSP to process quickly images collected from the real world using well trained AI models ({\textbf{For D1}}). However, the collected data could still burden the network. For example, the authors in \cite{wang2022semantic} state that a pair of sensing devices can generate 3.072 megabytes of data per second, which challenges conventional communication systems. Fortunately, semantic communication \cite{yang2022semantic} is introduced to filter out irrelevant information from edge devices to reduce information redundancy of VSPs by extracting semantic data from raw data and expressing desired meanings ({\textbf{For D2}}). With the help of semantic communications, massive amounts of data can be circulated in the virtual transportation networks to empower Metaverse services. 

However, the introduction of semantic communication brings about higher requirements of the data security. Efficient semantic data sharing should be achieved between unknown VSPs and edge devices deployed in untrusted environments. However, the extracted semantic data can be tampered to almost the same descriptors (semantic similarities) but different desired meanings. The attacker (edge device) can modify the pixels of a sunflower to make it similar to the extracted semantic data (snowy mountain) in terms of semantic similarities but visually dissimilar \cite{du2022rethinking}, which affects the output of the AIGC models and is difficult for VSPs to detect the difference between the adversarial and authentic semantic data. Moreover, it is hard to detect and prevent semantic data mutations for virtual transportation networks in Metaverse. Since VSPs and edge devices are distributed, it is difficult to record, trace, and verify data transformations. To solve the aforementioned problems, the blockchain and Zero-Knowledge Proof techniques can be used ({\textbf{For D3}}). Thus, we present a blockchain-aided semantic communication framework with AIGC services to achieve virtual transportation networks in Metaverse. Here, blockchain-aided semantic communication can facilitate data circulation and economic activities between VSPs and edge devices in a decentralized manner \cite{lin2022unified}. Targeted semantic attacks \cite{du2022rethinking} are utilized to generate adversarial semantic data to improve their semantic similarities almost up to that of the extracted semantic data by training various loss functions. Zero-Knowledge Proof (ZKP) \cite{song2022traceable} is integrated into the proposed framework to process semantic data and securely guarantee correct transformations. The contributions of this paper are summarized as follows: 

\begin{itemize}
    \item We propose a blockchain-aided semantic communication framework for AIGC in virtual transportation networks that ensures the authenticity of semantic data transmitted from edge devices to VSPs to facilitate interactions of the physical and virtual domains.
    \item We illustrate how a training-based targeted semantic attack scheme generates adversarial semantic data (images) without revealing the authentic semantic data. The attack semantic data has almost the same semantic similarities as the authentic ones but is visually dissimilar to them.
    \item We, for the first time, design a blockchain and zero-knowledge proof-based semantic defense scheme to assure the authentication of semantic data. The scheme can utilize zero-knowledge proof to record the transformations of semantic data, and use blockchain to track and verify semantic data mutations.
\end{itemize}

The remainder of the paper is described as follows. Section \ref{sec_related} reviews previous works on secure semantic communication. Section \ref{sec_framework} demonstrates a blockchain-aided semantic communication framework for AIGC in Metaverse. Section \ref{sec_attack} illustrates a training-based targeted semantic attack scheme. Section \ref{sec_defense} designs a blockchain and zero-knowledge proof-based semantic defense scheme. The proposed mechanisms are evaluated in Section \ref{sec_exp}. Section \ref{sec_conclude} concludes the paper and elaborates the future work.

\section{Related Work}
\label{sec_related}

In this paper, we consider the integration of blockchain-aided semantic communications for Metaverse, which involves multiple emerging technologies. Therefore, we divide the related work into three parts: Blockchain-aided Semantic Communications, Semantic Attacks, and Semantic Defenses.

\subsection{Blockchain-aided Semantic Communications} 

Semantic communication \cite{yang2022semantic} can lighten virtual transportation network burdens by transmitting relevant semantic data to VSPs after processing original data by AI technologies in edge devices. Z. Weng \textit{et al.} \cite{weng2021semantic} utilized deep learning (DL) to identify the essential speech information with higher weights for semantic communication in dynamic channel conditions. To enable edge devices to perform DL-based semantic communication tasks, H. Xie \textit{et al.} \cite{xie2020lite} proposed a lite-distributed semantic communication framework for edge devices to transmit low-complexity texts by optimizing training processes. 

Although semantic communication can help Metaverse reduce information redundancy, it can not handle the challenge that how to construct trust among unknown edge devices and VSPs to facilitate data sharing and economic activities. Blockchain \cite{nakamoto2008bitcoin} is a peer-to-peer network that can construct decentralized ledgers for participants to share data. The integration of blockchain and AI-based semantic communication can empower Metaverse ecosystems to carry out rich activities between the physical and virtual domains \cite{yang2022fusing}. Y. Lin \textit{et al.} \cite{lin2022unified} proposed a unified blockchain-semantic framework to enable Web 3.0 services to implement on-chain and off-chain interactions. A proof of semantic mechanism is proposed to verify semantic data before adding it to blockchain. However, they do not mention the performance indicators of semantic data. Y. Lin \textit{et al.} \cite{lin2022blockchain} proposed a blockchain-based semantic exchange framework that can mint Non-Fungible Tokens (NFT) for semantic data, utilize the game theory to facilitate exchange, and introduce ZKP to enable privacy-preserving. However, they do not consider semantic attacks that may reduce exchange efficiency.

\subsection{Semantic Attacks and Defenses}

The introduction of semantic communication brings about security issues for Metaverse. However, current research on the security of semantic communication is still in its infancy. Q. Hu \textit{et al.} \cite{hu2022robust} analyzed semantic noise that causes semantic data to express misleading meanings. To reduce the effects caused by semantic noise, they added weight perturbation to adversarial training processes, suppressed noise-related and task-unrelated features, and designed semantic similarity-based loss functions to reduce transmitting overheads. X. Luo \textit{et al.} \cite{luo2022encrypted} focused on privacy leakages when sharing background knowledge. They introduced symmetric encryption to adversarial training processes to encrypt and decrypt semantic data to ensure confidentiality. H. Du \textit{et al.} \cite{du2022rethinking} focused on the semantic Internet-of Things (SIoT) and proposed new performance indicators for SIoT to quantify security issues, including semantic secrecy outage probability and detection failure probability. They focused on image transmission-oriented semantic communication and divided semantic attacks into targeted and untargeted semantic attacks. The targeted attack can generate adversarial semantic data (images) that can be recovered to a given target requested by receivers. The adversarial semantic data has almost the same descriptors (semantic similarities), but is visually dissimilar \cite{tolias2019targeted}. The untargeted semantic attack can generate adversarial semantic data that minimizes semantic similarities. They do not pursue to be recovered to any target by receivers.

In this paper, we study the targeted semantic attacks and introduce ZKP to differ in semantic similarities between the adversarial and target images to protect semantic data. ZKP has been widely used in blockchain to enable privacy-preserving and authenticity. R. Song \textit{et al.} \cite{song2022traceable} utilized NFT and ZKP to construct a traceable data exchange scheme in blockchain, which can protect data privacy and exchange fairness. Z. Wang \textit{et al.} \cite{wan2022zk} designed a ZKP-based off-chain data feed scheme to take in off-chain sensitive data to execute the business logic of smart contract-based applications. H. Galal \textit{et al.} \cite{galal2022aegis} leveraged ZKP and smart contracts to hide details of NFTs and swap NFTs in a fair manner. Y. Fang \textit{et al.} \cite{fan2023validating} utilized ZKP to verify the authenticity of model prediction processes without leasing private parameters of deep learning models. However, the above methods do not consider how to use ZKP to prevent targeted semantic attacks to detect adversarial semantic data.

\subsection{Artificial Intelligence Generated Content}
AIGC refers to the use of AI algorithms, natural language processing (NLP), and computer version (CV) methods to produce a variety of media forms, including text, images, and audio. In terms of text content generation, an epoch-making AIGC application is the ChatGPT, a conversational language model developed by OpenAI~\cite{pavlik2023collaborating}. ChatGPT is a type of the Generative Pre-trained Transformer (GPT) model that is trained on a huge amount of conversational data. ChatGPT is able to generate text that sounds as if it were written by a human. This makes it helpful for a variety of purposes, including chatbots, virtual assistants, and language translation. For image generation, diffusion model~\cite{croitoru2022diffusion}, a new class of state-of-the-art generative models, have demonstrated exceptional performance in Image Generation tasks and have surpassed the performance of generative adversarial networks (GANs)~\cite{creswell2018generative} in numerous tasks. Stable Diffusion, a AIGC model that is released by Stability AI, is an open-source text-to-image generator that creates amazing artwork in a matter of seconds. It operates with unprecedented speed and quality on consumer-grade GPUs. Furthermore, AIGC techniques continues to advance in the field of audio generation. The authors in~\cite{shih2022theme} propose a deep learning-based method that employs contrastive representation learning and clustering to automatically derive thematic information from music pieces in the training data. The simulation results show that the proposed model is capable of generating polyphonic pop piano music with repetition and plausible variations.

One of the primary benefits of AIGC is that it can be produced at a number and speed that would be difficult for human beings to do alone. It also permits a great degree of consistency and precision. Therefore, the development of AIGC technologies can provide a strong boost to the evolution of the Internet. Notably, despite the potential benefits, there are also some worries about the ramifications of AIGC, including the possibility for prejudice, the semantic errors in the generated content, and the blurring of the boundary between human- and machine-generated content. Therefore, it is significant to ensure the correctness of the AIGC, avoiding attacks from malicious parties that causes resource waste to affect the quality of AIGC services.

\section{Blockchain-aided Semantic Communication Framework for AIGC in Metaverse}
\label{sec_framework}

\begin{figure*}[!t]
  \centering
  \includegraphics[width=0.9\textwidth]{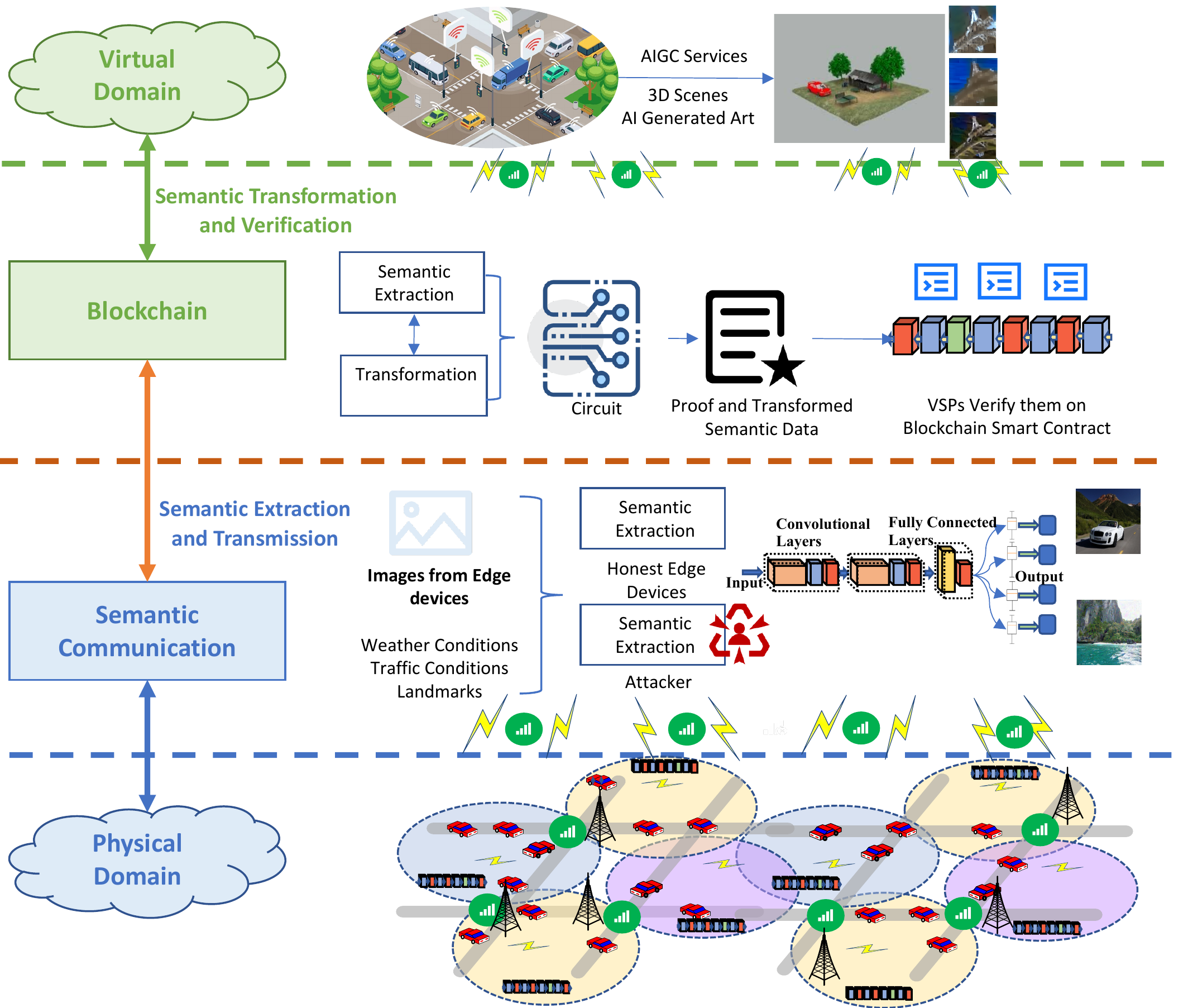}
  \caption{Blockchain-aided Semantic Communication Framework for AIGC in Metaverse}
  \label{fig_framework}
\end{figure*}

The implementation of virtual transportation networks requires the following main steps: 1) Semantic extraction and transmission for data interactions between physical and virtual domains, 2) Semantic transformation and verification, and 3) AIGC for Metaverse, as shown in Fig. \ref{fig_framework}.

\subsection{Semantic Extraction and Transmission}
Pedestrians are in danger when auto-driving models in vehicles are not trained well enough, which motivates the development of virtual transportation networks in the Metaverse. Therefore, it is necessary to digitize physical domains using data produced in the real world to simulate environments for training vehicles and drivers. VSPs can take pictures using edge devices like smartphones, cameras, and sensors in physical domains to obtain information about the weather, traffic, and geographical landmarks that can be used to train and test the detection systems of vehicles in virtual domains. Virtual domain output can be fed back into physical domains to configure vehicles for better and safer performance. Based on the simulated environment, inexperienced drivers and vehicles can practice their reactions under unfamiliar weather or traffic conditions in Metaverse in a safe way. Therefore, VSPs need to frequently interact with edge devices supported by a tremendous amount of data to construct virtual transportation networks in Metaverse to provide services, which challenges the transmission capabilities of edge devices and VSPs.

Unfortunately, conventional communication systems cannot afford such frequent interactions between the physical and virtual domains, which will cause virtual transportation networks in Metaverse to lack enough data to simulate virtual environments. Semantic communication, a completely new paradigm that extracts semantic meaning from raw data for transmission, can be utilized to resolve this challenge. Instead of transmitting original data, edge devices can extract the semantic data and transmit to VSPs. The VSPs can then use the received semantic data to generate simulation environments for drivers and autonomous vehicle training on the virtual road. For instance, edge devices, e.g., smartphones, positioned at various locations along the same road may gather photographs of traffic conditions from their perspective. To reduce information redundancy, they can utilize semantic segmentation modules to crop key components of images as semantic data~\cite{ng2022stochastic}. Then edge devices only transmit semantic data to VSPs to report traffic conditions on roads.

However, since VSPs have to collect semantic data from multiple edge devices to train virtual transportation networks in different situations, malicious edge devices could falsify semantic data (images) and corrupt the training process, which is dangerous for pedestrians and drivers. In this paper, we study the targeted semantic attack \cite{du2022rethinking,tolias2019targeted} in that the adversarial and authentic semantic data have almost the same semantic similarities but are totally irrelevant. The semantic similarities are calculated with the help of high-dimensional descriptors extracted by a convolutional neural network (CNN). However, malicious edge devices could train a corresponding neural network to modify some pixels in attack images to achieve similar descriptors as the authentic images to corrupt the construction of virtual transportation networks. The training-based targeted semantic attack scheme is illustrated in Section \ref{sec_attack}.


\subsection{AIGC in Metaverse}

After receiving semantic data (images) from edge devices deployed in different locations, VSPs can perceive conditions or views of landmarks, and render images by AIGC services in Metaverse. For example, semantic data of landmarks from different perspectives can be utilized by VSPs to render 3D scenes to provide users with seamless experiences. VSPs can also utilize views of landmarks to generate artworks or avatars in Metaverse. Therefore, AIGC services play an important role in virtual transportation networks to facilitate the use of data resources and the applications of Metaverse. Besides, semantic data is important for subsequent AIGC services since its quality may affect the content generated by AIGC.

However, since semantic data circulated in the Metaverse may be corrupted by the aforementioned targeted semantic attacks produced by malicious edge devices, the AIGC services may do useless work and provide users with hateful content. For example, VSPs want to collect images of famous landmarks (i.e., Eiffel Tower) while malicious edge devices may extract unrelated images of flowers to corrupt the subsequent AIGC services that renders 3D scenes with different perspectives of the landmark. VSPs also want to perceive images of landmarks to generate artworks or avatars while attackers transmit irrelative semantic data of animals to obstacle AIGC services. Since the adversarial semantic data (images) modified by attackers have almost the same semantic similarities, it is difficult and time-consuming for VSPs to verify the authenticity of images. Therefore, it is necessary to design a mechanism to ensure the security of data transmission to protect the security of AIGC services.

\subsection{Semantic Transformation and Verification}
Malicious semantic data can affect the security of virtual transportation services in Metaverse. Modifying pixels in unrelated semantic data increases the semantic similarity score, causing the VSPs to use the wrong semantic data as input to AIGC and corrupt the virtual environment. Inspired by \cite{tolias2019targeted}, image transformations can be performed to distinguish semantic similarities between the adversarial and authentic images. However, malicious edge devices can continue adjusting pixels in irrelative images to make them similar to transformed semantic data. Moreover, it is impossible for edge devices to transform semantic data unlimited times. A possible and practical solution is to record and verify transformations performed on semantic data.

Therefore, we propose a blockchain and zero-knowledge proof-based semantic defense scheme in Section \ref{sec_defense}. The logic of transformations is recorded on the circuit produced by the zero-knowledge algorithm. Edge devices utilize extracted semantic data as inputs to generate proof of transformations and output transformed semantic data. They send the proof and semantic data to VSPs for verification. Since VSPs and edge devices are distributed in unknown Metaverse environments represented by avatars, the blockchain should be used to record and verify transformations to prevent data mutations.

\section{Training-based Targeted Semantic Attack Scheme}
\label{sec_attack}
Although the blockchain-aided semantic communication framework for AIGC in Metaverse can reduce information redundancy and establish decentralized trust between unknown edge devices and VSPs, malicious edge devices may conduct training-based targeted semantic attacks to corrupt semantic data (images) circulated in the Metaverse services. The targeted semantic attacks refer to transmitting adversarial semantic data with almost the same semantic descriptors but visually dissimilar to the authentic one \cite{du2022rethinking}, which is difficult for VSPs to utilize semantic similarities (inner product of descriptors among images) evaluation to distinguish them. In this section, we illustrate the workflow of the training-based targeted semantic attack targeted semantic attacks \cite{du2022rethinking}\cite{tolias2019targeted}. 

\textbf{Descriptor Extraction.} Since extracted semantic data (images) is difficult to evaluate, edge devices can utilize a CNN to map images to high dimensional descriptors. Then VSPs can use the descriptors to distinguish semantic similarities of images. The process of descriptor extraction is illustrated as follows.

\begin{figure}[!t]
  \centering
  \includegraphics[width=3.5in]{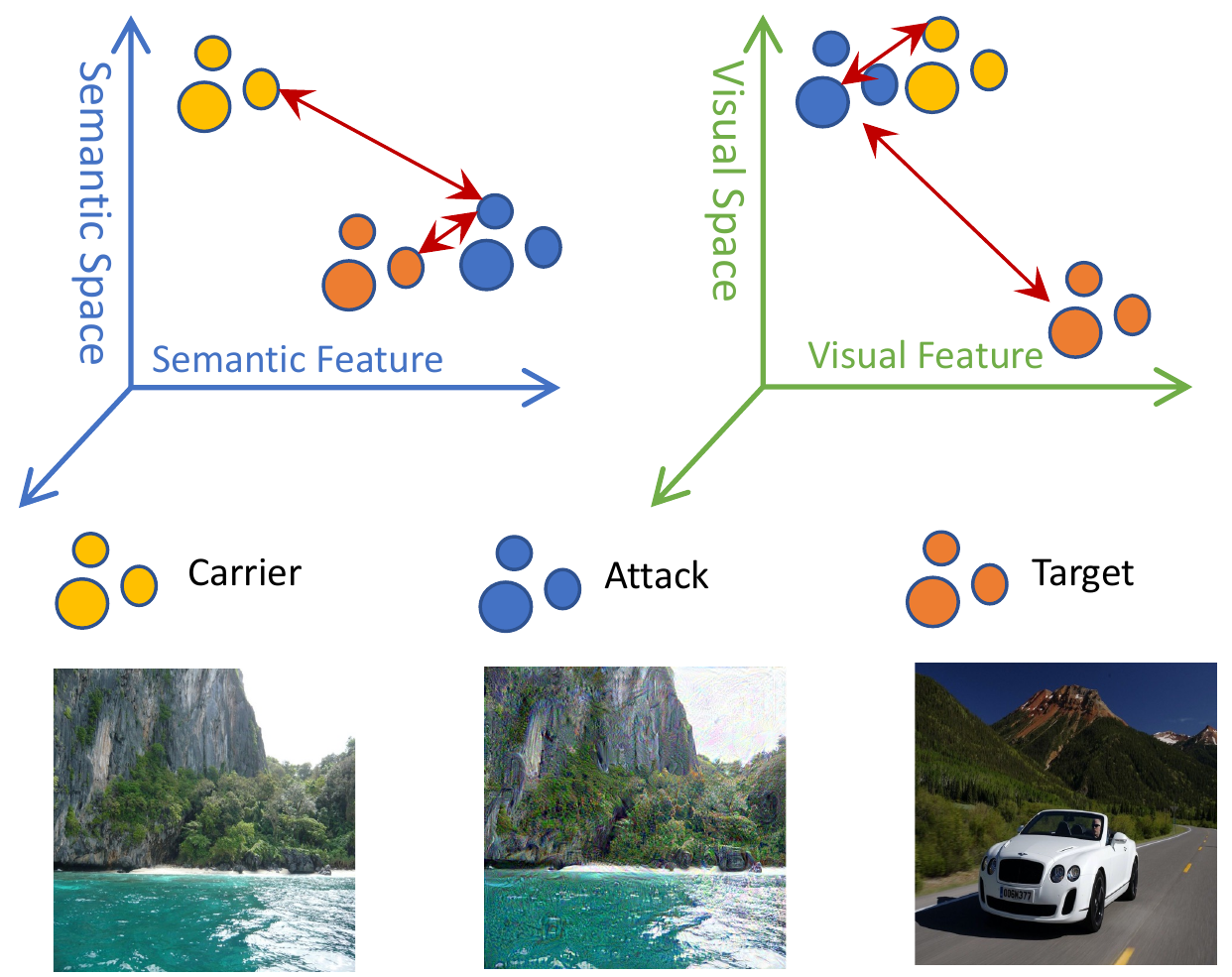}
  \caption{Relationship among Images}
  \label{fig_attack_exp}
\end{figure}

Let us denote three types of semantic data produced by malicious edge devices, including the adversarial semantic data $\mathbf{x_a}$, the authentic one $\mathbf{x_t}$, and the carrier one $\mathbf{x_c}$. The authentic semantic data $\mathbf{x_t}$ is extracted from original images by semantic extraction in edge devices according to requirements and associated with label $y_t=f_{\mathbf{c}}(x_t)$, which has semantic similarities with $\mathbf{x_a}$. $f_{\mathbf{c}}(\cdot)$ is utilized to classify semantic data. The carrier semantic data $\mathbf{x_c}$ is an auxiliary image to help malicious edge devices generate $\mathbf{x_a}$, which is classified as $y_c=f_{\mathbf{c}}(x_c) \neq y_t$ and has visual similarities with $\mathbf{x_a}$. The adversarial semantic data $\mathbf{x_a}$ is produced by training networks to learn how to modify pixels, which can achieve that $\mathbf{x_a}$ has almost the same descriptors but is visually dissimilar from the authentic one $\mathbf{x_t}$ generated by semantic extraction, as shown in Fig. \ref{fig_attack_exp}. Besides, $\mathbf{x_a}$ is visually similar to $\mathbf{x_c}$ but classified incorrectly as $y_t$. Therefore, the descriptor extraction process is vital for malicious edge devices to produce adversarial images $\mathbf{x_a}$.

The input image $\mathbf{x}$ should be re-sampled to $\mathbf{x}^s$ with the same dimension $s$ to make $\mathbf{x_t}$ and $\mathbf{x_c}$ with the same resolution. The image $\mathbf{x}^s$ is utilized as an input to train a Fully CNN given by $\mathbf{g_{x^s}}=g(\mathbf{x}^s):\mathbb{R}^{W \times H \times 3} \rightarrow \mathbb{R}^{w \times h \times d}$ to implement feature extraction where $W \times H \times 3$ and $w \times h \times d$ are weight, height, and channel of images. A pooling layer is utilized to map the input tensor $\mathbf{g_{x^s}}$ and the descriptor $\mathbf{h_{x^s}}=h(\mathbf{g_{x^s}})$ by the CNN with network parameters $\mathbf{\theta}$, which is mapped by $h: \mathbb{R}^{w \times h \times d} \rightarrow \mathbb{R}^d$. The descriptor is the output of the pooling layer, which can be utilized to compare with other descriptors to calculate semantic similarities. The $l_2$ normalization is utilized to easily compare semantic similarities. 

\textbf{Loss Function.} Considering the goal of malicious edge devices is to make $\mathbf{x_a}$ almost the same as $\mathbf{x_t}$ in terms of semantics but visually similar to $\mathbf{x_c}$, the loss function consists of the performance loss $l_{\mbox{ts}}(\mathbf{x},\mathbf{x}_t)$ and the distortion loss between $\mathbf{x}$ and $\mathbf{x}_c$, which can be defined as

\begin{equation}
    L_{\mbox{ts}}(\mathbf{x}_c,\mathbf{x}_t ; \mathbf{x})=l_{\mbox{ts}}(\mathbf{x},\mathbf{x}_t) + \lambda \Vert \mathbf{x} - \mathbf{x}_c \Vert^2,
\end{equation} where $\lambda$ is a hyper-parameter to show the impact of the distortion loss.

\textbf{Performance Loss.} Let us assume that malicious edge devices can access to the network structure of the descriptor extraction \cite{tolias2019targeted} since it is necessary for edge devices to know the evaluation standard of VSPs. Considering different scenarios where targeted semantic attacks generate, referring to \cite{tolias2019targeted}, we introduce the three empirical forms of the performance loss as follows.

\textit{Global Descriptor Loss} is suitable that malicious edge devices even know all parameters of the network of the descriptor extraction. Thus, the performance loss function $l_{\mbox{ts}}$ can be given by calculating the inner product of $\mathbf{x_a}$ and $\mathbf{x_t}$ as follows:

\begin{equation}
    l_{\mbox{global}}(\mathbf{x},\mathbf{x}_t) = 1 - \mathbf{h}^{\top}_{\mathbf{x}} \mathbf{h}_{\mathbf{x}_t}.
\end{equation}

\textit{Activation Tensor Loss} is adapted to the scenario where the outputs of the network of the descriptor extraction are the same for $\mathbf{x_a}$ and $\mathbf{x_t}$ before down-sampling to resolution $s$, which can be denoted by the mean squared difference of $\mathbf{g}_{\mathbf{x}}$ and $\mathbf{g}_{\mathbf{x}_t}$ as follows:

\begin{equation}
    l_{\mbox{tensor}}(\mathbf{x},\mathbf{x}_t) = \frac{\Vert \mathbf{g}_{\mathbf{x}} - \mathbf{g}_{\mathbf{x}_t} \Vert^2}{w \cdot h \cdot d}.
\end{equation}

\textit{Activation Histogram Loss} is utilized to preserve first-order statistics of activations $\Vert u(\mathbf{g_x},\mathbf{b})_i$ per channel $i$ to achieve identical extracted descriptors regardless of spatial information in images, which can be denoted as follows:

\begin{equation}
    l_{\mbox{hist}}(\mathbf{x},\mathbf{x}_t) = \frac{1}{d} \sum_{i=1}^d \Vert u(\mathbf{g_x},\mathbf{b})_i - u(\mathbf{g_{x_t}},\mathbf{b})_i \Vert,
\end{equation} where $\mathbf{b}$ is the histogram bin centers.

\textbf{Optimization.} Our goal is to find the optimal loss function that minimizes the semantic similarities of $\mathbf{x_t}$ and $\mathbf{x}$, which equivalently optimizes network parameters $\mathbf{\theta}$. We can use the Adam algorithm to update $\mathbf{\theta}$ as

\begin{equation}
    \theta_{t+1} = \theta_{t} - \eta \frac{\rho_t}{\sqrt{\upsilon_t + \epsilon}},
\end{equation} where $\eta$ is the learning rate, $\rho_t$ and $\upsilon_t$ are the first-order and second-order momenta of gradients, and $\epsilon$ is utilized to prevent the $\sqrt{\upsilon_t + \epsilon}$ from being zero. Therefore, the adversarial semantic data $\mathbf{x}_a$ can be expressed by 

\begin{equation}
    \mathbf{x}_a = \mathop{\arg\min}_{\mathbf{x}} L_{\mbox{ts}}(\mathbf{x}_c,\mathbf{x}_t ; \mathbf{x}),
\end{equation} where $L_{\mbox{ts}}$ can be replaced by $l_{\mbox{global}}$, $l_{\mbox{tensor}}$, or $l_{\mbox{hist}}$ according to different scenarios. 

\begin{figure*}[!t]
  \centering
  \includegraphics[width=0.9\textwidth]{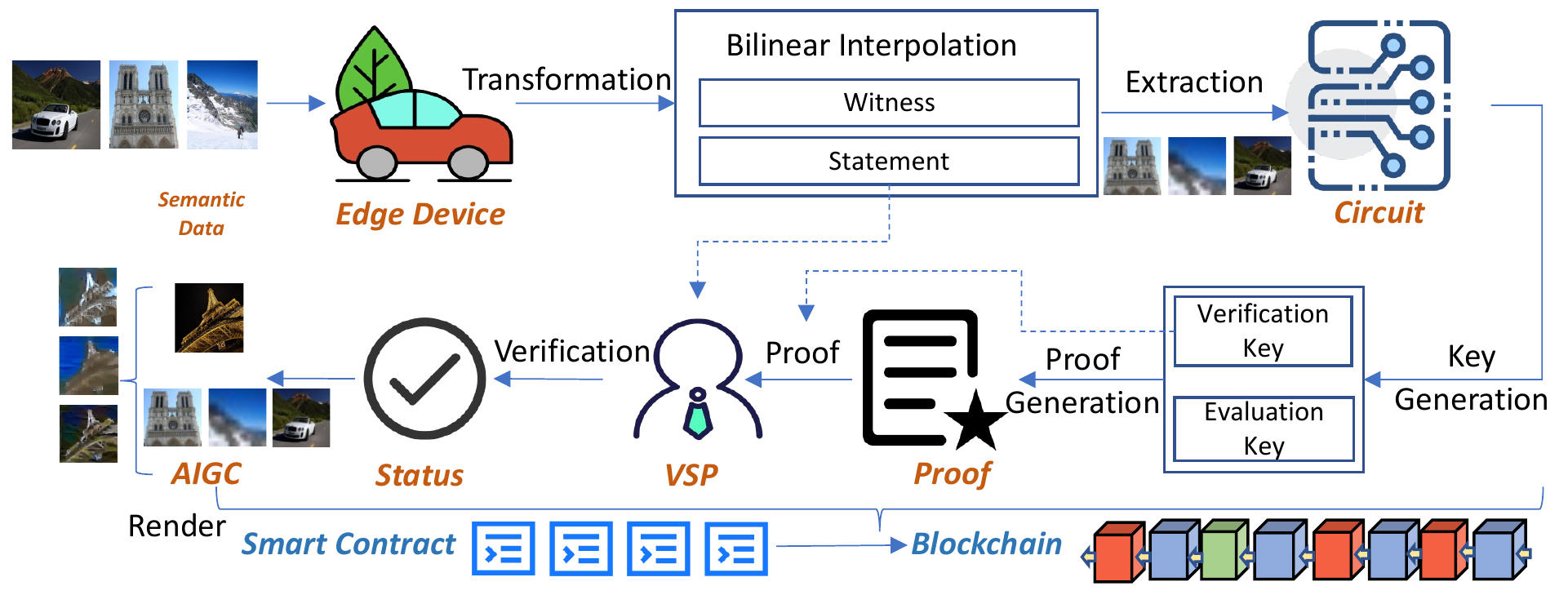}
  \caption{Blockchain and Zero-Knowledge Proof-based Semantic Defense Scheme}
  \label{fig_zkp}
\end{figure*}

\section{Blockchain and Zero-Knowledge Proof-based Semantic Defense Scheme}
\label{sec_defense}

Since the adversarial semantic data generated by malicious edge devices has almost the same descriptors (semantic similarity) but different desired meanings from the authentic ones produced by honest edge devices, inspired by \cite{du2022rethinking}\cite{tolias2019targeted}, we utilize blockchain and zero-knowledge proof-based semantic defense scheme to help VSPs to identify attack images transmitted in the Metaverse. Instead of submitting extracted semantic data directly, edge devices should transform or process semantic data by the bilinear interpolation algorithm \cite{castleman1996digital}, and utilize Zero-Knowledge Proof to record and verify transformations. The details of the proposed scheme are elaborated as follows, as shown in Fig. \ref{fig_zkp}.

\textbf{Transformation.} The transformation of semantic data is a training-free defense method that uses visual invariance to distinguish adversarial and authentic semantic extraction. The reason is that attackers adjust some pixels to make descriptors of adversarial images similar to the authentic ones \cite{du2022rethinking}. As a result, we attempt to increase visual invariance by blurring extracted images using the spatial transformation of the bilinear interpolation algorithm.

Let us assume that $(x_1,y_1)$, $(x_1,y_2)$, $(x_2,y_1)$, and $(x_2,y_2)$ are four points in the extracted images. Then the targeted points $(x,y)$ can be obtained by the spatial transformation, i.e., bilinear interpolation in the $x$ and $y$ directions. The spatial transformation can be considered a mapping function $f(\cdot,\cdot)$. Thus, the linear interpolation in the $x$ direction can be derived as follows:

\begin{equation}
    f(x,y_1) \approx \frac{x_2-x}{x_2-x_1}f(x_1,y_1)+\frac{x-x_1}{x_2-x_1}f(x_2,y_1)
\end{equation}
and
\begin{equation}
    f(x,y_2) \approx \frac{x_2-x}{x_2-x_1}f(x_1,y_2)+\frac{x-x_1}{x_2-x_1}f(x_2,y_2).
\end{equation}
The targeted points are transformed by the linear interpolation in the $y$ direction as follows:

\begin{footnotesize}
\begin{equation}
\begin{split}
    & f(x,y) \approx \frac{y_2-y}{y_2-y_1}f(x,y_1)+\frac{y-y_1}{y_2-y_1}f(x,y_2) \\
    & \approx \frac{(x_2-x)(y_2-y)}{(x_2-x_1)(y_2-y_1)}f(x_1,y_1)+\frac{(x-x_1)(y_2-y)}{(x_2-x_1)(y_2-y_1))}f(x_2,y_1)\\
    &+\frac{(x_2-x)(y-y_1)}{(x_2-x_1)(y_2-y_1)}f(x_1,y_2)+\frac{(x-x_1)(y-y_1)}{(x_2-x_1)(y_2-y_1)}f(x_2,y_2).
\end{split}
\end{equation}
\end{footnotesize}

Although edge devices can perform transformations to obtain the blurred semantic data that can distinguish from the adversarial one, it is difficult for VSPs to verify whether the blurred semantic data is derived from authentic transformations. Malicious edge devices may modify some pixels to make descriptors of their semantic data close to that of honest edge devices after transformations. Therefore, to assure the authenticity of the blur transformation for semantic data, we utilize ZKP to record transformations and blockchain to verify them. The proposed mechanism is a tuple of 3 polynomial-time schemes after extracting a circuit mapping the transformation, including Key Generation, Proof, and Verification, which works as follows:

\textbf{Extraction.} The mechanism initiates the logic in a computation circuit $\mathsf{C}$ using the simple arithmetic expression mapping the transformation $f$ to construct the public statement $\mathsf{s}$ and the private witness $\mathbf{w}$ \cite{fan2023validating}. The circuit implements the logic of bilinear interpolation via circom \cite{circom}, a ZKP circuit compiler. The circuit can provide a relation between the inputs and outputs semantic data which can be used for verifiable computation on transformations without disclosing the inputs. The process of extraction can be illustrated as follows:

\begin{equation}
    \mathsf{Extract}(f) \rightarrow (\mathsf{s},\mathsf{w}).
\end{equation}

\textbf{Key Generation.} Edge devices take a security parameter $1^{\lambda}$ and the transformation circuit $\mathsf{C}$ as inputs to generate a common reference string $\mathsf{crs}$. The common reference string $\mathsf{crs}$ includes an evaluation key $\mathsf{crs.ek}$ and a verification key $\mathsf{crs.vk}$ for proof and verification. The process of key generation can be expressed as follows:

\begin{equation}
    \mathsf{KeyGen}(1^{\lambda},\mathsf{C}) \xrightarrow{\mathsf{C}} \mathsf{crs(ek,vk)}.
\end{equation}

\textbf{Proof.} Edge devices take the evaluation key $\mathsf{crs.ek}$, the statement $\mathsf{s}$ and the witness $\mathsf{w}$ related to the transformed and original semantic data, which satisfies $\mathsf{C(s,w)}=1$. The statement $\mathsf{s}$ and the witness $\mathsf{w}$ stand for the public and private information corresponding to the transformation relation. Thus, a zero-knowledge proof $\pi$ is generated to reflect and verify the relation. The process of proof generation can be denoted as follows: 

\begin{equation}
    \mathsf{Prove}(\mathsf{crs.ek,s,w}) \xrightarrow{\mathsf{C}} \pi.
\end{equation}

\textbf{Verification.} VSPs utilize smart contracts deployed on blockchain to verify the authenticity of transformations and implement the business logic of blockchain-aided semantic communication for AIGC in Metaverse. Since the verification process is implemented in the blockchain, edge devices and VSPs can query verification results to construct trust in a decentralized manner. The verification key $\mathsf{crs.vk}$, the statement $\mathsf{s}$, and the proof $\pi$ are the inputs written into smart contracts to determine whether to accept or reject the proof according to outputs. When the status of the output is 1, the verification succeeds and VSPs accept the semantic data provided by edge devices; otherwise, VSPs refuse to accept the proof corresponding to semantic data produced by malicious edge devices. The process of proof verification can be given as follows:

\begin{equation}
    \mathsf{Verify}(\mathsf{crs.vk,s},\pi) \rightarrow \{0,1\}.
\end{equation}

The semantic defense scheme should satisfy the following properties including completeness, soundness, and zero-knowledge \cite{song2022traceable}\cite{wan2022zk}\cite{galal2022aegis}\cite{fan2023validating}.

\textbf{Completeness} means that an honest edge device with a valid witness $\mathsf{w}$ can convince an honest VSP for the authenticity of transformed semantic data generated from the circuit $\mathsf{C}$. If the transformed semantic data extracted by bilinear interpolation is correct and edge devices construct the proof $\pi$ correctly through the proof circuit $\mathsf{C}$ and the evaluation key $\mathsf{crs.ek}$, then the VSPs can use the verification key $\mathsf{crs.vk}$ generated by the proof circuit $\mathsf{C}$, the proof $\pi$, and the public information (statement) $\mathsf{s}$ to obtain a verification result. For each $\mathsf{C(s,w)}=1$, the proof $\pi$ generated from an honest edge device will be accepted with probability $1$, which can be denoted as follows:

\begin{equation}
    \Pr \left[
    \begin{array}{c}
      \mathsf{KeyGen}(1^{\lambda},\mathsf{C}) \rightarrow \mathsf{crs(ek,vk)} \\
      \mathsf{Prove}(\mathsf{crs.ek,s,w}) \rightarrow \pi \\ 
      \mathsf{Verify}(\mathsf{crs.vk,s},\pi) = 1 
    \end{array}
\right] = 1.
\end{equation}

\begin{figure}
\centerline{\includegraphics[width=0.45\textwidth]{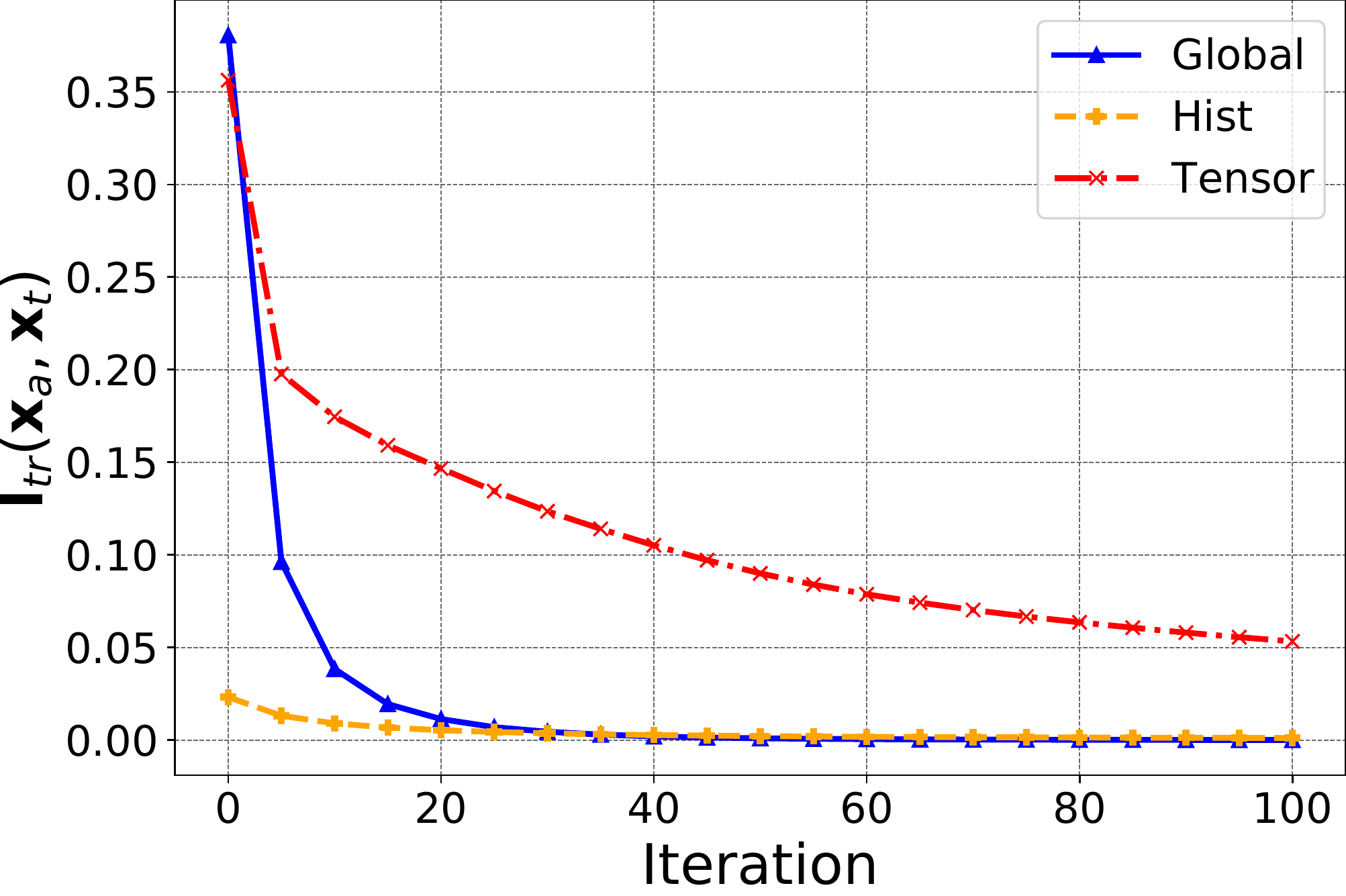}}
\caption{Performance Loss of Attacks\label{fig_loss_perf}}
\end{figure}

\begin{figure}[!t]
    \centering      
    \subfigure[Semantic similarity between the adversarial image and the authentic]{\includegraphics[width=1.6in]{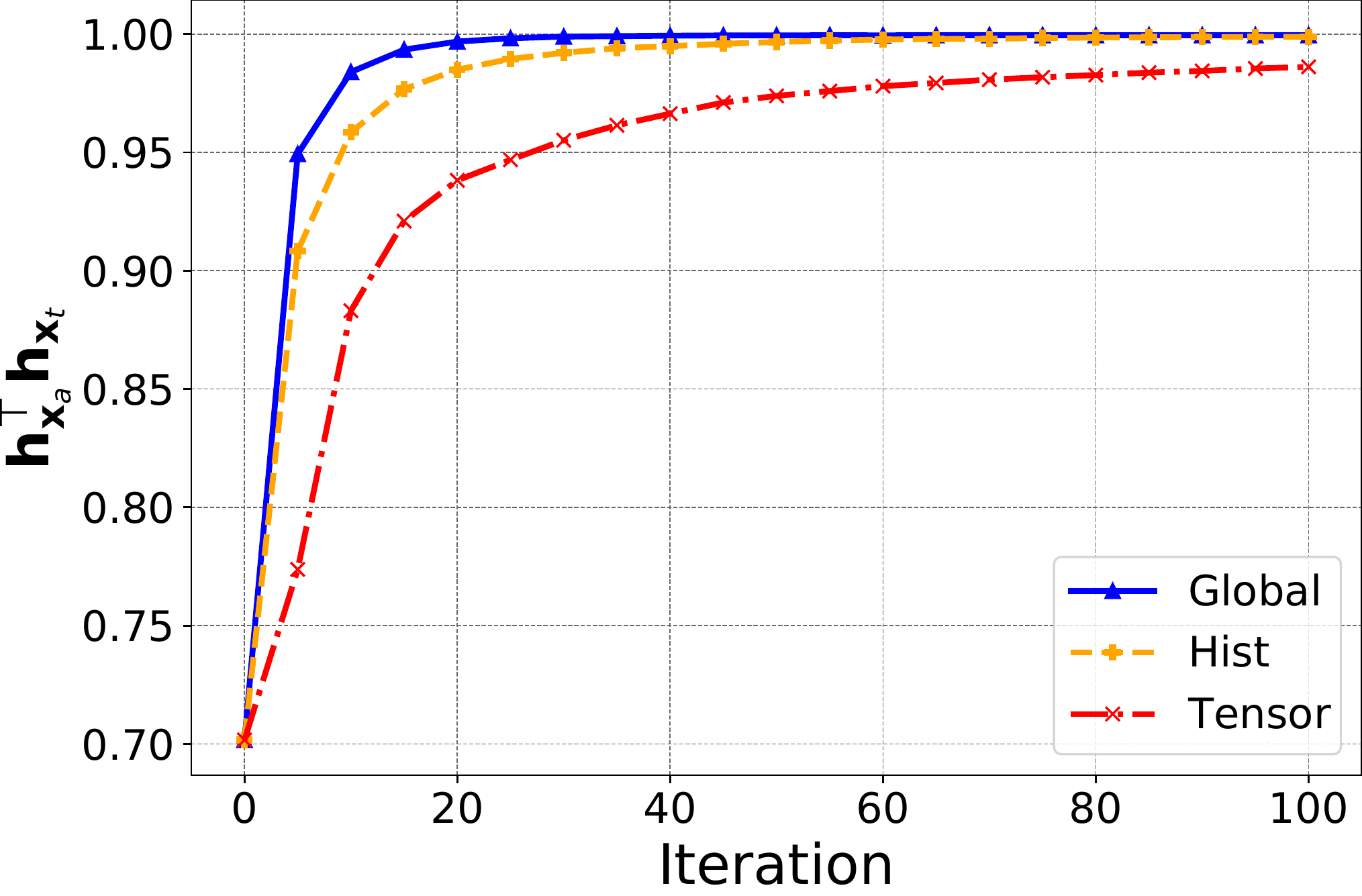}}
    \subfigure[Semantic similarity between the adversarial image and the carrier]{\includegraphics[width=1.6in]{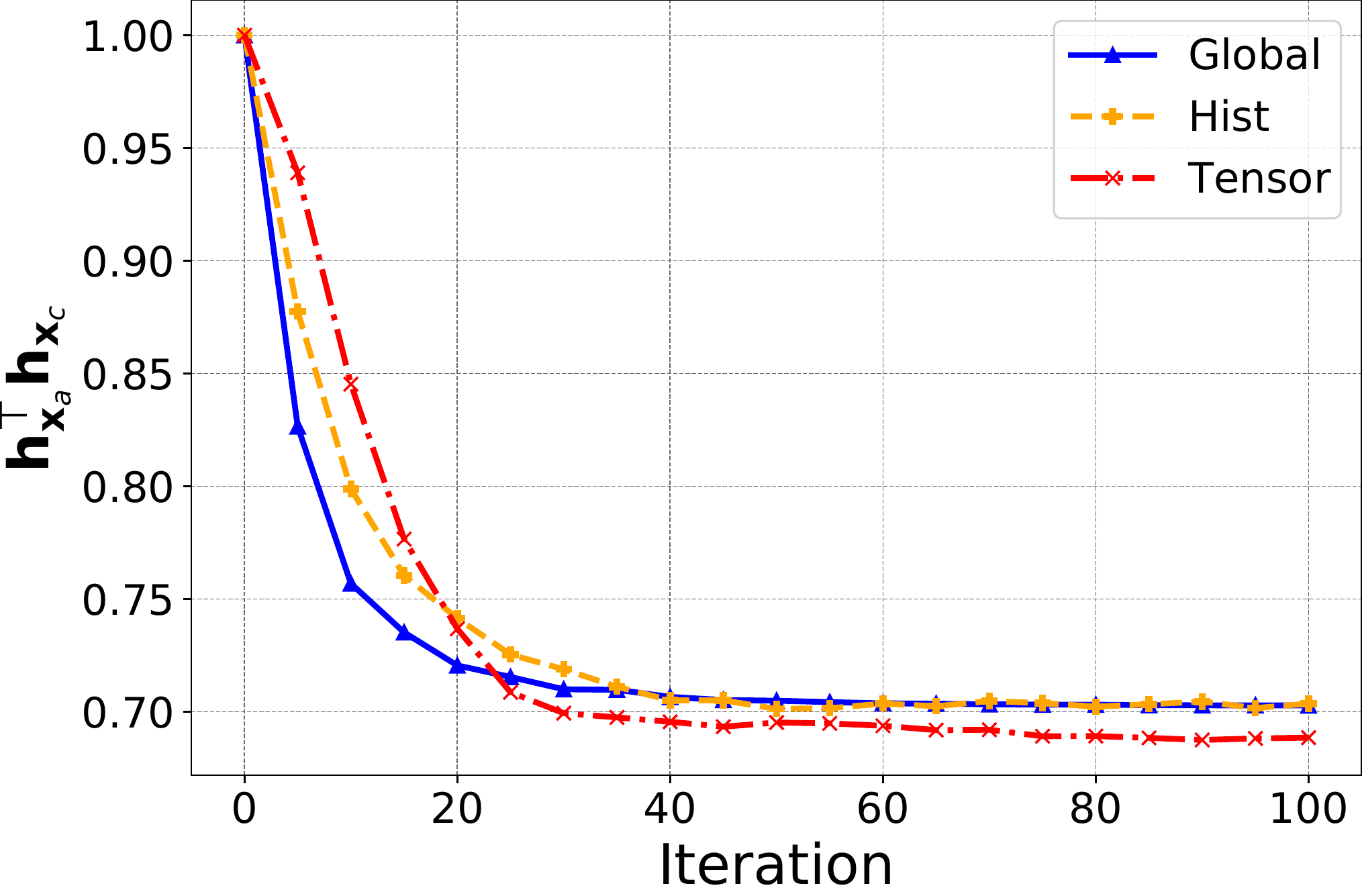}}
    \caption{Semantic Similarity Performance of Attacks}
    \label{fig_similarity_attack}
\end{figure}

\textbf{Soundness} denotes that malicious edge devices cannot provide VSPs with authentic transformed semantic data, which means that edge devices can provide valid witnesses if they can produce valid proofs. If $(\mathsf{s,w})$ is not a valid input to the proof circuit $\mathsf{C(s,w)}=0$, there is a polynomial time extractor $\mathsf{Ext}$ for a malicious edge device $\mathcal{A}$ that makes the probabilities of $\mathsf{Verify}(\mathsf{crs.vk,s},\pi^*)=1$ are negligible. 
The adversarial advantage of soundness $\mathsf{Adv}_{\mathcal{A}}^{sd}(\lambda)$ can be represented as follows: 

\begin{equation}
    \Pr \left[
    \begin{array}{c}
      \mathsf{KeyGen}(1^{\lambda},\mathsf{C}) \rightarrow \mathsf{crs(ek,vk)} \\
      \mathcal{A}(\mathsf{crs.ek,s}) \rightarrow \pi^* \\ 
      \mathsf{Ext(crs.ek,s,\pi^*)} \rightarrow \mathsf{w} \\
      \mathsf{Verify}(\mathsf{crs.vk,s},\pi^*) = 1 
    \end{array}
\right] \leq \mathsf{negl}(\lambda).
    \label{eq_soundness}
\end{equation}

Malicious edge devices $\mathcal{A}$ can hardly falsify semantic data or make honest VSPs $\mathcal{V}$ accept invalid proofs about transformations $f$. According to (\ref{eq_soundness}), 
VSPs cannot accept false proof $\pi$ from $\mathcal{A}$ except for a negligible probability $\mathsf{negl}(\lambda)$.

\textbf{Zero-knowledge} represents that VSPs can only verify the authenticity of transformed semantic data while they cannot obtain the original semantic data unless edge devices send it to them. Let us assume that there are probabilistic polynomial time simulators $\mathsf{S_1}$, $\mathsf{S_2}$, and malicious edge devices $\mathcal{A}_1$, $\mathcal{A}_2$. The simulator $\mathsf{S_1}$ can produce a common reference string that is used by the simulator $\mathsf{S_2}$ to generate a simulated proof. Then the proposed mechanism has the zero-knowledge property if the adversarial advantage of zero-knowledge $\mathsf{Adv}_{\mathcal{A}}^{zk}(\lambda)$ satisfies:

\begin{equation}
    |\Pr(\mathsf{real}) - \Pr(\mathsf{sim})| \leq \mathsf{negl}(\lambda),
    \label{eq_zero_knowledge}
\end{equation} where $\Pr(\mathsf{real})$ and $\Pr(\mathsf{sim})$ can be denoted as

\begin{equation}
    \Pr(\mathsf{real}) = \Pr \left[
    \begin{array}{c}
        \mathsf{KeyGen}(1^{\lambda},\mathsf{C}) \rightarrow \mathsf{crs} \\
        \mathcal{A}_1(f) \rightarrow (\mathsf{s,w}) \\
        \mathsf{Prove(crs,s,w)} \rightarrow \pi \\
        \mathcal{A}_2(\mathsf{crs,s,w,\pi}) = 1     
    \end{array}
\right] 
\label{zero_analys_1}
\end{equation}

and

\begin{equation}
    \Pr(\mathsf{sim}) = \Pr \left[
    \begin{array}{c}
        \mathsf{S_1}(1^{\lambda},\mathsf{C}) \rightarrow \mathsf{crs} \\
        \mathcal{A}_1(f) \rightarrow (\mathsf{s,w}) \\
        \mathsf{S_2(crs,s)} \rightarrow \pi \\
        \mathcal{A}_2(\mathsf{crs,s,w,\pi}) = 1     
    \end{array}
\right].
\label{zero_analys_2}
\end{equation}

\begin{figure*}[!t]
  \centering
  \includegraphics[width=7in]{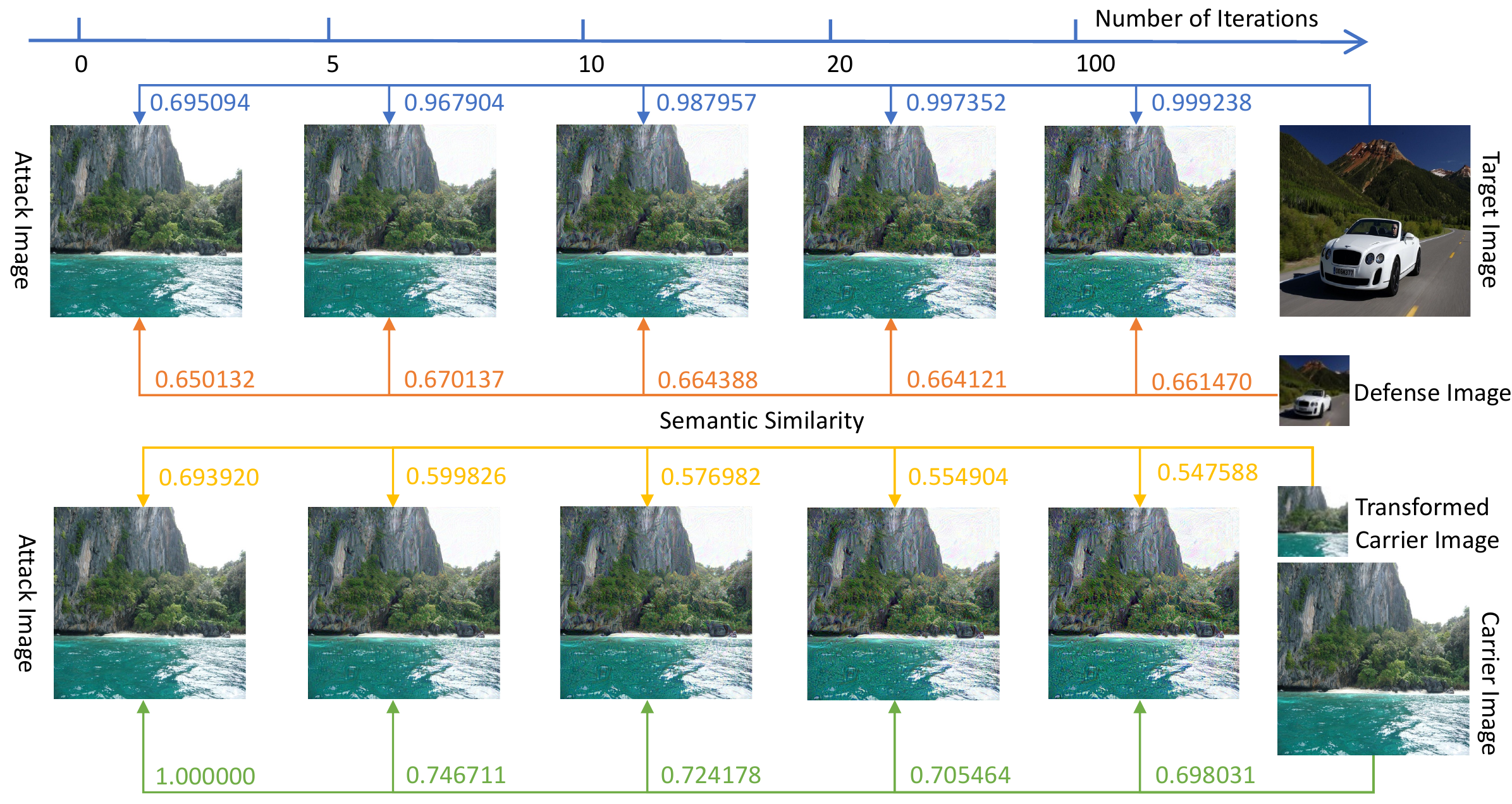}
  \caption{Semantic Similarity Comparison Between Attack and Defense Schemes}
  \label{fig_attack}
\end{figure*}

\begin{figure}[!t]
    \centering      
    \subfigure[Semantic similarity between the adversarial image and the authentic]{\includegraphics[width=1.6in]{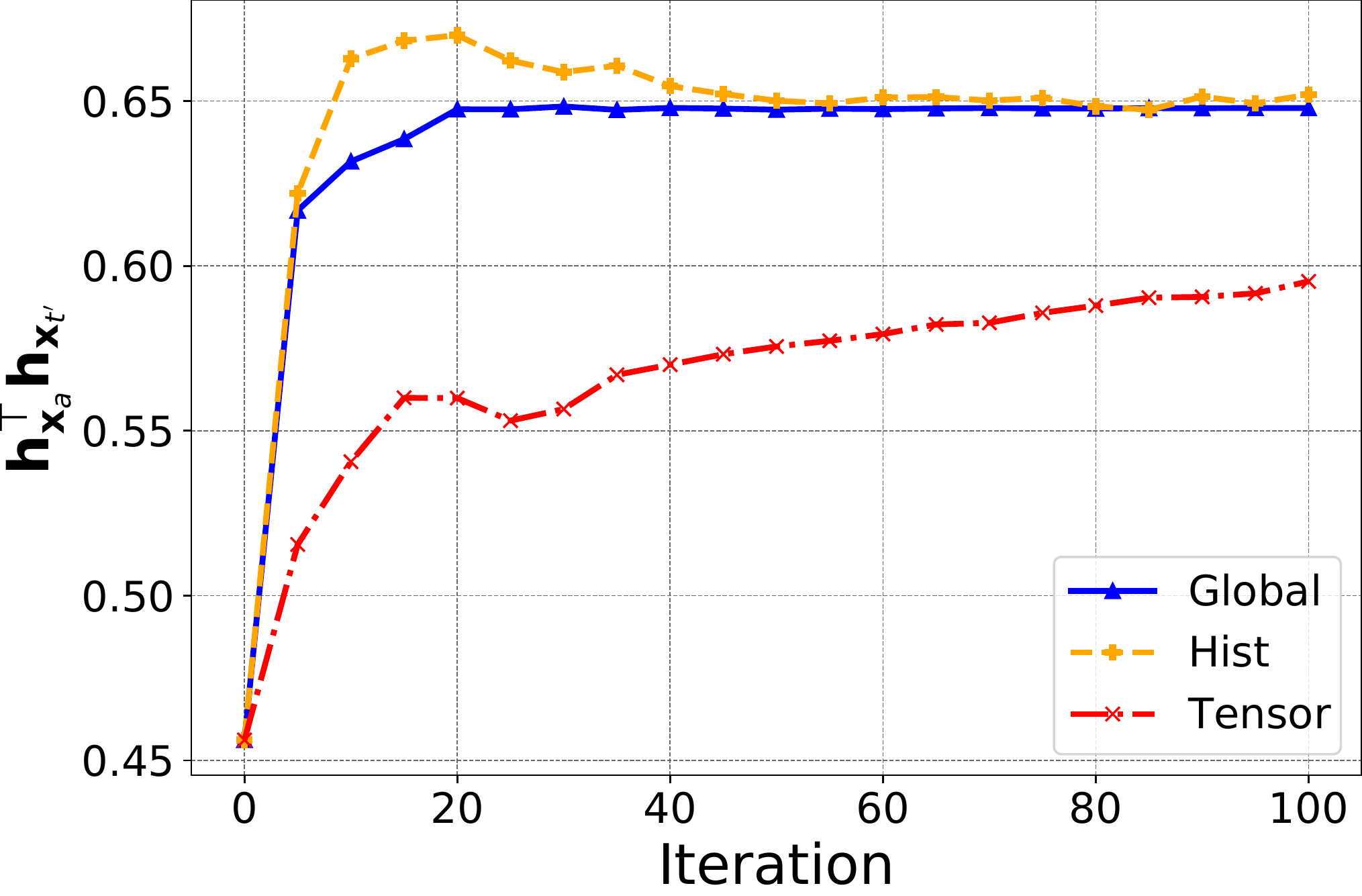}}
    \subfigure[Semantic similarity between the adversarial image and the carrier]{\includegraphics[width=1.6in]{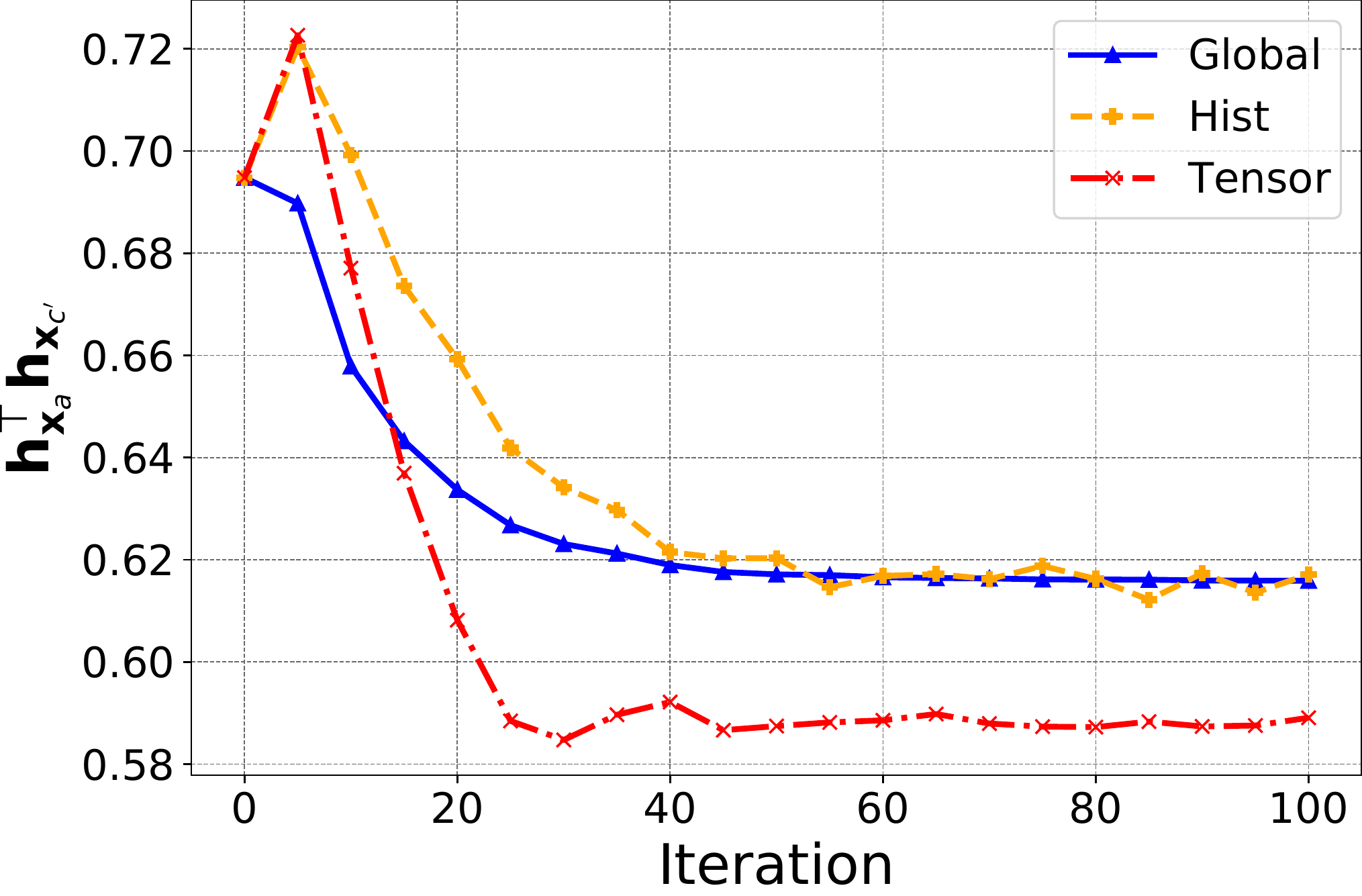}}
    \caption{Semantic Similarity Performance of Defenses}
    \label{fig_similarity_defense}
\end{figure}

VSPs who know public statements mapping with semantic data can hardly learn any knowledge about semantic data before and after transformation. According to (16), VSPs cannot know anything about semantic data other than what can be inferred from transformation.

\section{Experimental Evaluations}
\label{sec_exp}

We simulate the proposed mechanism on Ubuntu 20.04LTS, Intel Xeon, 8 core, 64G memory, and 25000Mb/s with 2 Tesla V100, Go 1.19.4, Node v14.17.0, PyTorch 1.12.1, Torchvision 0.13.1, and CUDA 10.2. We perform experiments on a standard image benchmark, Revisited Paris \cite{radenovic2018revisiting}, to measure the attack and defense performance among edge devices and VSPs. We exploit pre-trained networks on ImageNet \cite{russakovsky2015imagenet} and AlexNet \cite{krizhevsky2017imagenet} to execute the attacks and defenses with 100 iterations. Unless otherwise mentioned, we use these parameters referring to \cite{tolias2019targeted}. The learning rate $\eta$ is set to 0.01. The hyper-parameter $\lambda$ that adjusts the impact of distortion loss is 0. The training process performs 100 iterations for our experiments.

\subsection{Performance of Semantic Attack Scheme}\ 

Fig. \ref{fig_loss_perf} is the performance loss of the proposed semantic attack scheme with different loss functions [Global, Hist, Tensor] corresponding to the global descriptor, activation tensor, and activation histogram as the number of iterations increases. Tensor converges much slower than Global and Hist, which requires more iterations to reach its plateau.

Fig. \ref{fig_similarity_attack} is the semantic similarity attack performance between the adversarial semantic data and the authentic or the carrier one with three loss functions [Global, Hist, Tensor]. As shown in Fig. \ref{fig_similarity_attack} (a) and Fig. \ref{fig_similarity_attack} (b), Global and Hist converge around the 40th iteration which is much faster than Tensor. The semantic similarity performance of attacks is consistent with Fig. \ref{fig_loss_perf} which shows the same trend in three loss functions.

\subsection{Performance of Semantic Defense Scheme}\

Fig. \ref{fig_attack} is the semantic similarity comparison between attack and defense schemes by displaying example images. Fig. \ref{fig_similarity_defense} is the semantic similarity (descriptors) defense performance between the adversarial semantic data and the authentic one or the carrier one with three loss functions [Global, Hist, Tensor]. Compared with Fig. \ref{fig_similarity_attack}, the adversarial image (semantic data) can differ in semantic similarities when applying the proposed semantic defense scheme. Since the extracted semantic data is required to execute the defense scheme and the execution can be assured by zero-knowledge proof, the images can differ in descriptors. As shown in Fig. \ref{fig_similarity_attack} (a) and Fig. \ref{fig_similarity_defense} (a), the semantic similarity can reduce by up to 35\% between the adversarial and the authentic images. Fig. \ref{fig_similarity_attack} (b) and Fig. \ref{fig_similarity_defense} (b) show that the semantic similarity can decrease 10\% between the adversarial and the carrier images. 

\begin{figure}[!t]
    \centering      
    \subfigure[Consensus]{\includegraphics[width=1.6in]{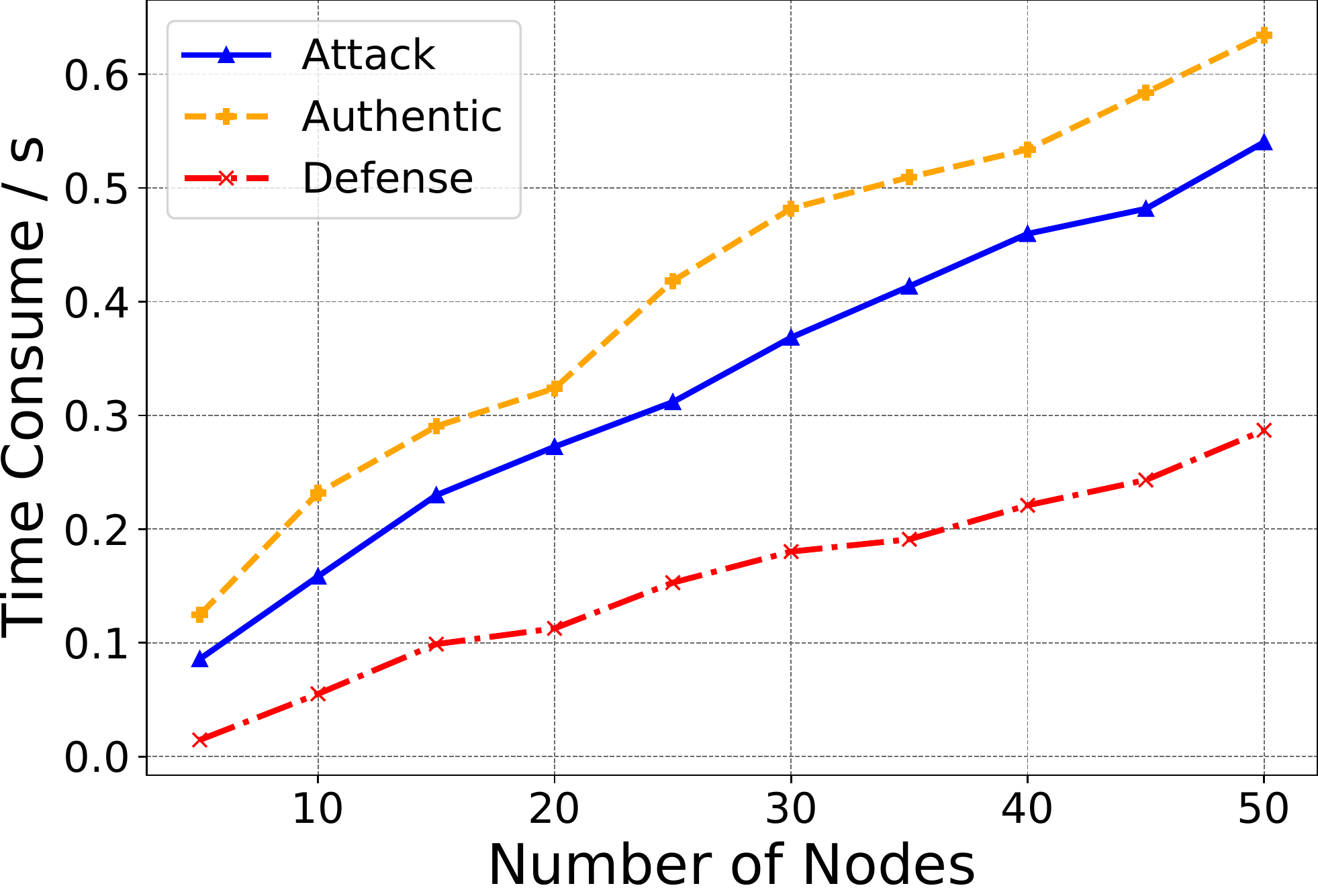}}
    \subfigure[ZKP Computation]{\includegraphics[width=1.4in]{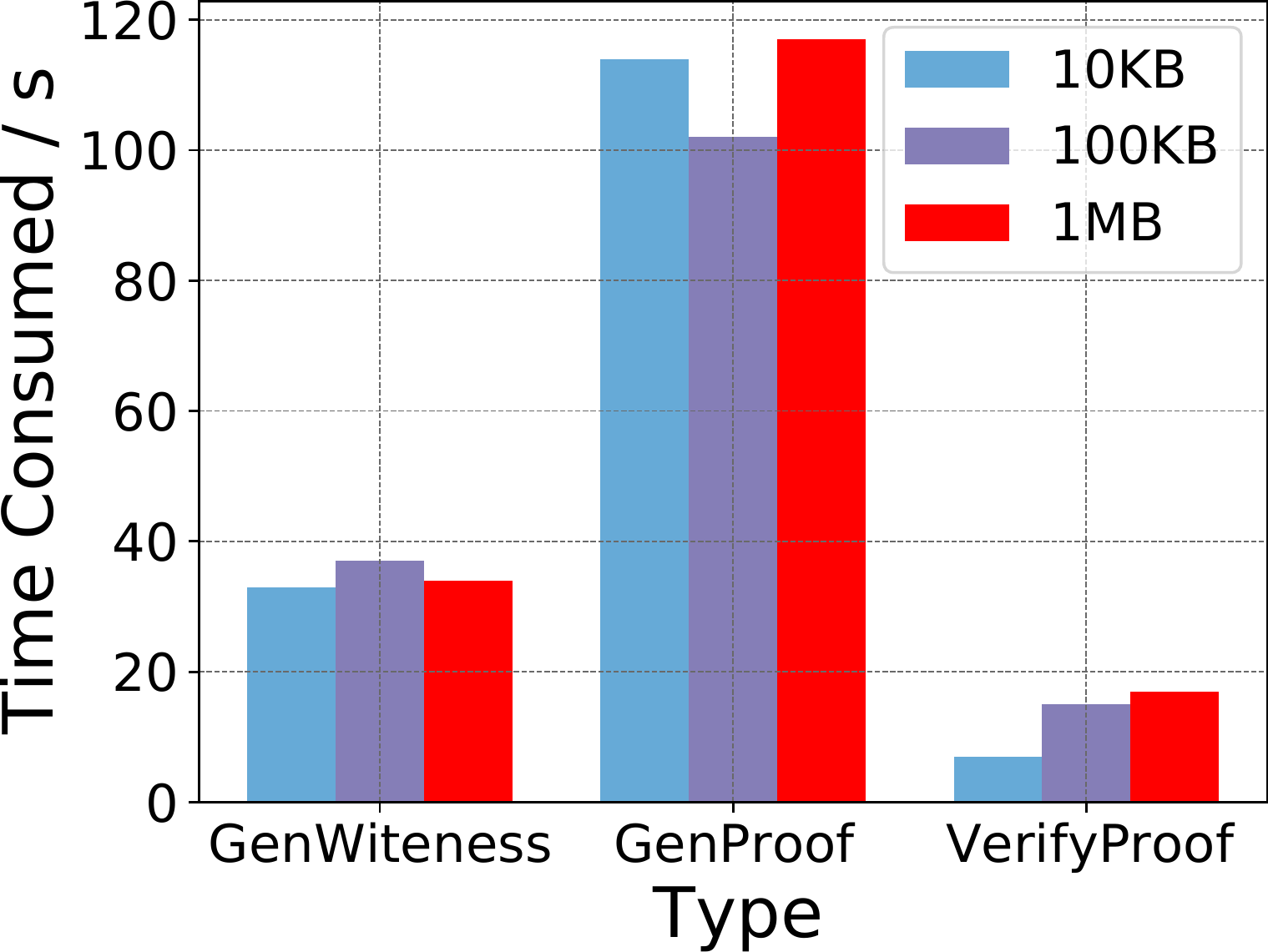}}
    \caption{Consensus and Computation Overheads}
    \label{fig_blockchain}
\end{figure}

Fig. \ref{fig_blockchain} (a) is the blockchain consensus overhead for [Attack, Authentic, Defense] types of semantic data (image) with [275 KB,467 KB,10 KB] data sizes when the number of nodes is [5,50,5]. As shown in Fig. \ref{fig_blockchain} (a), the proposed scheme can reduce the time consumed in blockchain to improve consensus efficiency due to the attack and defense schemes cropping authentic semantic data while differing in semantic similarities according to Fig. \ref{fig_similarity_defense}. Fig. \ref{fig_blockchain} (b) is the ZKP computation overhead for [GenWitness, GenProof, VerifyProof] operations with [10KB,100KB,1MB] data sizes. We can see from Fig. \ref{fig_blockchain} (b) that the ZKP computation overhead depends on the GenProof operation, while the sizes of semantic data affect little on the proposed scheme. Besides, the computation overhead for verifying proofs is less than for generating proofs, which can be written into smart contracts \cite{song2022traceable}\cite{wan2022zk}\cite{galal2022aegis}.

\section{Conclusion and Future Work}
\label{sec_conclude}
In this paper, we first integrated blockchain-aided semantic communication into Metaverse to support AIGC services in virtual transportation networks. We also presented a training-based targeted semantic attack scheme to illustrate potential attacks by generating adversarial semantic data with the same descriptors but different desired meanings. To prevent the above attack, we designed a blockchain and zero-knowledge proof-based semantic defense scheme that records transformations of semantic data and verifies mutations in a decentralized manner. Simulation results show that the proposed mechanism can differ in the descriptors between the adversarial semantic data and the authentic one. The defense scheme can identify malicious edge devices falsifying semantic data to corrupt AIGC services in virtual transportation networks. In future research work, we will study how to mitigate malicious edge devices and facilitate resource allocation with economic incentive mechanisms in the proposed framework.

\bibliographystyle{IEEEtran}
\bibliography{ref}



\end{document}